# Metrics for BPEL Process Reusability Analysis in a Workflow System

A. Khoshkbarforoushha [a,b], P. Jamshidi [c], M. Fahmideh [d], L. Wang [e], R. Ranjan [b]

[a] School of Computer Science, Australian National University, Canberra, Australia

[b] CSIRO Computational Informatics, Building 108, Australian National University, Canberra, Australia

[c] Lero - The Irish Software Engineering Research Centre, School of Computing, Dublin City University, Dublin, Ireland

[d] Automated Software Engineering Research Group, Electrical and Computer Engineering Faculty, Shahid Beheshti University, Tehran, Iran

[e] School of Computer Science, China University of Geosciences, Wuhan 430074, Hubei, PR China

Corresponding author: Lizhe Wang, email: Lizhe.wang@gmail.com

**Abstract**

*This work proposes a quantitative metric to analyze potential reusability of a BPEL (Business Process Execution Language) Process. The approach is based on Description and Logic Mismatch Probability of a BPEL Process that will be reused within potential contexts. The mismatch probabilities have been consolidated to a metric formula for quantifying the probability of potential reuse of BPEL processes.*
*An initial empirical evaluation suggests that the proposed metric properly predict potential reusability of BPEL processes. According to the experiment, there exists a significant statistical correlation between results of the metric and the experts' judgments. This indicates a predictive dependency between the proposed metric and potential reusability of BPEL processes as a measuring stick for this phenomena. If future studies ascertain these findings by replicating this experiment, the practical implications of such a metric are early detection of the design flaws and aiding architects to judge various design alternatives.*

## 1. Introduction

Web service composition is an emerging approach to integrate applications and produce intra-organizational business processes. In Service-Oriented Computing (SOC), a business process is a coarse-grained composite web service executing a control-flow (i.e. service logic) to meet a business goal. Among various technologies, BPEL (Business Process Execution Language) is a de facto standard that is utilized to realize required orchestration and choreography between diverse web services. In fact, BPEL is a workflow-oriented composition model that brings a central piece in the heavily modularized SOC model. Throughout this paper composite service and BPEL process terms are used interchangeably.

It has been widely recognized that reusability is inherent to service-oriented solutions and studies demonstrate that organizations regard reuse as a top driver for Service-Oriented Architecture (SOA) adoption [1]. In fact, reusability is acknowledged as the main purpose in the design of services [2].

According to research work reported in the literature [3][4] there is a strong demand for reusing process models, in particular BPEL processes. In this regard, researchers end up in different reuse strategies including process template [5][6], reference process [7][8], ad hoc modifications to existing process models, reusing parts of business processes with the aid of BPEL Fragments [3], and recently by applying Service Supervision Patterns[9].

A BPEL process would not be reused in potential contexts if there were any mismatches between potential requirements and those the service expected to realize. The context is categorized to current solution that a service being reused and potential future solutions that it might be reused. Therefore, service designers should keep in mind that any service they produce can potentially become a reusable asset [10]. Process designers should not exclusively focus on requirements of the initial consumers of a service, but rather should adopt appropriate reuse strategies and also undertake more extensive business analysis in order to determine more general requirements to produce reusable assets. Taking this approach into account, software developers should support in their preparation of software for potential reuse [11]. In this regard, Frakes and Kang [12] emphasize that there is a clear need for future research to identify good ways to estimate the number of potential reuses. Therefore, Potential Reusability ($R_p$) analysis refers to predicting to what extent a software element can be reused in the future [13]. This means, in this paper we do not aim to measure BPEL process reusability. Instead, this paper is to propose a measure to predict the potential reusability of BPEL processes as service-centric implemented processes. This is called predictive validity[14].

Although there is a set of guidelines and approaches [15][16] that help architects to ensure a balance of proper logic encapsulation and adherence to standard description in software services, there is no quantitative metric to analyze the extent to which a designed BPEL process can be reused in the future. Previous researches on service reusability, while contributing to the field, have been subject to the following criticisms. These include being interface-driven and neglecting logical constructs, being too complex and hard to collect. In contrast, our metrics not only consider logical activities within BPEL processes, their calculations are based on BPEL process standard specifications, hence all the computations are automated and do not require user intervention.

The key contributions of the paper are definition and empirical validation of a quantitative metric to analyze potential reusability of BPEL processes. The proposed approach is based on Description and Logic Mismatch Probability estimation of composite web services that will be reused across potential solutions. In order to validate the proposed metric, the reusability principles as theoretical properties were investigated. Thereafter, a data set consisting 70 professional BPEL processes was collected and an evaluation workshop was then held to provide empirical basis in order to demonstrate and calculate their feasibility, validation, and confidence interval respectively.

## 2. Related work

Software reuse has been the subject of extensive investigation in the traditional software development paradigms such as Object-Oriented Development (OOD) and Component-Based Development (CBD) [17]. In fact, reusability has been accounted as the main reason for introducing object-oriented approach to software development [18]. Many researchers developed significant metrics for measuring reusability [19][20][21][22]. As a next step toward increasing reusability, CBD approach has emerged in which software components as fundamental reusable building blocks are adopted for software development. Nonetheless, lack of standard for component interoperability [23] made obstacle to improve reusability in the CBD approach.

Although, the previous efforts in OO and CBD have been applied to software reuse, in particular measurable metrics, there are significant challenges to accommodate these metrics in SOA context:

- **Different abstraction level**: Services are typically at higher level of abstraction and encapsulation than software components and classes [24]. OO metrics only measure reusability by considering source code analysis and class complexity. With the same manner, the CBD metrics merely consider a component, including a number of classes, while a BPEL process can be developed through adopting a number of software services. Perepletchikov et al. have evaluated the applicability of some of the well-established OO metrics [25] in the context of SOA, particularly BPEL-based services. To illustrate deficiency of applicability of existing OO metrics, a predefined software project has been developed through both OO and BPEL-based approach respectively. The quantitative comparison of results shows that the existing metrics cannot be applied and there is a need to tailor existing metrics or introduce new ones to SOA.

- **Different measurable aspects:** OO metrics measure reusability by considering the critical aspects of classes such as number of methods in a class, depth of inheritance hierarchy, number of children, lack of cohesion in methods, coupling between classes, and number of responsibilities provided by class [21]. These metrics are not applicable in BPEL context since BPEL language constructs are totally different, meaning that there is no relationship between BPEL language measurable factors and OO. Existing CBD metrics measure externally visible features of a component mainly interface methods [22] rather than internal logic of the component that is concerned in the context of BPEL services.

With respect to the aforementioned issues, we have to focus on the latest research work in reusability metrics in the SOA domain. In [26], authors suggest measuring reusability based on use of the service by service consumers. Specifically, the number of existing consumers of a service indicates the reusability of the service or service reuse index. Similarly, they defined the number of consumers of that operation across services and business processes as the extent of operation reusability or operation reuse index. Although their proposed metrics are technology-neutral and can also be applied to both atomic and composite services, they neglect a salient point. That is, their proposed metrics do not contemplate any features that helps architects to determine if the designed services can be reused less or more in the future. Apparently, current reusability of a service cannot be the only predictor of reusability.

In [27], authors presented a quality model for evaluating reusability of services. Their quantitative model is based on the key quality attributes of reusability including business commonality, modularity, adaptability, standard conformance, and discoverability that are derived from the key features of services in SOA. *Business Commonality* refers to the measurement of functionality and non-functionality degree of the service that is commonly used by the service consumers in a domain. *Modularity* measures the extent to which a service supplies distinct functionality without depending upon other services. *Adaptability* measures the capability of the service to be properly adapted to various service consumers. *Standard Conformance* measures the extent to which a service complies with the widely accepted industry standards. *Discoverability* measures the degree to which the service is simply and accurately found by consumers. As a matter of fact, the authors' work is more complement and comprehensive comparing to the work in[26], but it suffers from two shortcomings. Firstly, even though their metrics can be applied to both atomic and composite services, they do not consider the logic or control-flow of composite services. Secondly, none of their proposed metrics have been empirically validated.

There are also some metrics that measure cohesion and coupling of services through which reusability of a service can be inferred implicitly [28][29]. For instance, in [29] Perepletchikov et al. proposed a number of quantitative metrics for measuring cohesion of service. The proposed cohesion metrics can be accommodated during design time based on the operations exposed in service interface in term of *Service Interface Data Cohesion*, *Service Interface Usage Cohesion*, *Service Interface Implementation Cohesion* and *Service Interface Sequential Cohesion*. However, the research does not consider structures that most services are constructed via standard XML-based business process definition languages. In reality the approach is mainly limited to service consumer concerns going to be published and utilized in a broader range and provider concerns such as logic or control-flow of services have been neglected. Moreover, achieving a proper assessment of reusability of a service based on coupling or cohesion is not straightforward and needs to be demonstrated by various replications with different settings.

## 3. Metric rationales

### 3.1 Reusability in SOA context

BPEL process reusability is defined as the extent to which a BPEL process (i.e. composite service) can be reused in other contexts, organizations, or SOA solutions with minimal effort and change (minimal change will be specified in more details throughout the section 4) . Reusability in service-oriented area is not an isolated concept and architects should decide about that, while other contexts and projects are contemplated. This means, an architect must analyze whether the given composite service can be reused in other business processes or domains. The term context, in this paper, is categorized to current and potential ones. Reusability in the current and potential contexts makes sense if we consider the point that every identified, specified, realized, and finally implemented BPEL process should be reused in different possible service-oriented solutions. Other possible assumptions could be those cases where services are going to be published and utilized in a broader range and new business ventures in which they could take advantage of the reuse capabilities and reusable assets. According to this assumption, the service reusability must be calculated with respect to the current and potential reusability.

### 3.2 The impact of business rule changes on BPEL process reusability

By adopting the provided reusability definition, we aim to justify the rationales behind our metrics through proposing a research question that is the inspiring source of our metric research work: "*In which conditions a BPEL process cannot be reused?*" Apparently a service cannot be reused where it cannot be matched with new or potential requirements in other contexts or solutions in the future. Now, this question may arise: "*what are these requirements?*" These requirements are, in fact, business rules. To be more specific, we have to emphasize that even though they are not the same concept, they do have close relationship with each other. One of the distinction between them is that unlike requirements (particularly software requirement), business rules convey and enforce the policy of an organization [30] and they are usually flexible due to the nature of organizational change and often consist of very granular steps. Subsequently, business rules are often subject to frequent changes. On the other hand, the most important sources of requirement changes are business rules[31].

The smallest unit of business change is a business rule change. Regarding business rules drive the business process[32], business rules and business processes are interrelated, whenever the rules change, processes may change accordingly. Wagner [33] classified business rules in four basic types: integrity constraints, derivation rules, reaction rules, and deontic assignments rules.

According to the taxonomy of flexibility, there are five perspectives in business processes including functional, organizational, behavioral, informational, and operational. However, as specified in [34] the above-mentioned business rules primarily influence the informational and the behavioral perspectives of a business process.

The informational perspective defines the information that is exchanged between activities, whereas the behavioral perspective describes under which preconditions activities are executed. Integrity constraints specify preconditions for particular activity types in the business process model. For instance, "*After year 2012, patients without health card cannot place a request*" is an integrity constraint that primarily influence on the informational perspective. Derivation rules are statements that define new business concepts or facts based on the existing ones. For example, "*Old persons receives 15% discount for medical care*" is a derivation rule which make an impact on the informational perspective.

Reaction rules and deontic assignment rules state which activities need to be executed and in what order. Reaction rules define behavior in the first-person that participates in business process interactions, whereas deontic assignment rules specify control-flows of a process in terms of a third-person view point. These rules primarily influence the behavioral perspective of a business process. To illustrate the point, consider the *ManageEmergencyPatient* BPEL process. In some hospitals the sequence of emergency patient management is similar to the Fig. 1(a). That is, after receiving and processing client request, the patient is admitted and in the next step the payment verification scheme is launched. However, in the other hospitals it would be possible that the patient has to pay in advance in order to be provided with medical care and services, Fig. 1(b). Therefore, the given structure may mismatch with the business rule in some hospitals.

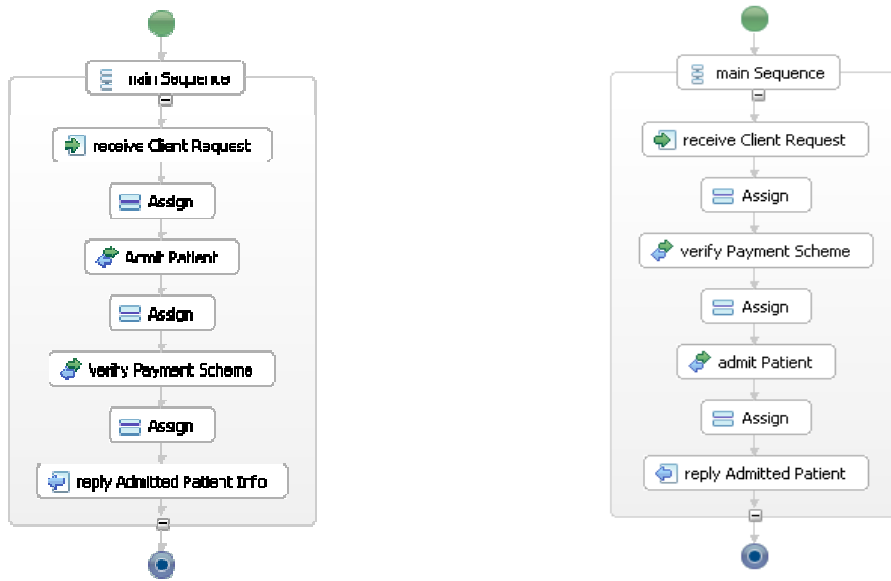

Fig. 1(a): First version of *ManageEmergencyPatient* BPEL process.　　　　Fig. 1(b): Second version of *ManageEmergencyPatient* BPEL process.

Services are realized at a higher-level of abstraction directly supporting business processes[35], on the condition that they are composed and orchestrated in accordance with business rules[36]. This means, the structure and arrangement of services within a composite one is imposed by business rules.

Therefore, any changes in the mentioned business rules lead to some changes in informational and behavioral perspectives of a business process. Since informational perspective refers to the information that is exchanged between the activities, so in case of any change, its main impact can be traced in web service interfaces. This is why every web service has an interface and the messages are exchanged through web service interfaces. If the change occurs on behavioral perspective, the orchestration of execution control-flow, which represents behavioral perspective of a business process, may change accordingly.

Any changes in business rules may provoke some kinds of mismatches for either interface or control-flow of a composite service in foreseen solutions. Consequently, a composite service cannot be reused in the future when it cannot be matched with the business rules of other contexts or solutions.

### 3.3 Basic concepts of potential reusability estimation

Frakes and Kang in [12] state that *"A key element in the success of reuse is the ability to predict needed variabilities in future assets."* In fact, predicting potential reusability is the function of estimating to what extent a BPEL process can be adapted with future requirements and business rules. Such adaptability and flexibility pave the way to judge potential reusability of a BPEL process. In other words, a BPEL process can be reused in the future if it can be matched with minimal effort to the expected one in foreseen contexts. Such a hypothesis is also supported on the basis of the research work such as [37] that proposes a process matching approach for flexible workflow process reuse.

To sum up, we formalize some definitions as follows:

**Definition 1:** (*Mismatch*). Mismatch refers to the point that a BPEL Process cannot be matched with other context requirements. In this regard, there are two kinds of mismatches including Description and Logic mismatch.

**Definition 2:** (*Description Mismatch*). Description mismatch denotes to the mismatch between the requirements and the description of the given composite service i.e. WSDL (Web Services Description Language).

When a service consumer either another service or an SOA architect want to use a given service based on their requirements, it may not be reused due to mismatch in WSDL including data types and messages of service operations. Apparently, some techniques could help an architect to utilize a service with mismatch in description; hence based on our reusability definition, some modifications would be tolerable. Therefore, description mismatch probability calculation of a composite service can provide new insight about its reusability.

**Definition 3:** (*Logic Mismatch*). Logic mismatch is defined as mismatch between the requirements and the logic of given composite service. Composite service logic is utilized through the control-flow of basic and structured activities within it. Regarding different contexts may enjoy dissimilar business rules, consequently, a service cannot be reused in potential contexts for the sake of disparate business rule which impose disparate structure and control-flow i.e. logic mismatch.

**Definition 4:** (*Mismatch Probability*). Mismatch probability (MMP) refers to the probability that a given composite service cannot be matched with other context requirements. Probability is a numerical measure of the likelihood of an event relative to a set of alternative events.

## 4. BPEL Process Reusability

In this section, the metrics including description and logic mismatch probability are introduced. To elaborate more on how the proposed metric works, after introducing each metric we will apply it to one of the important healthcare processes, Visit Preparation, as it is shown in Figure 2. This process was adopted from IBM SOA solution in healthcare organization[38]. To keep calculations simple, it is assumed that the given BPEL process interface has two operations (i.e. web methods). Table 1 contains their corresponding input and output parameters.

Table 1: operations of visit preparation process and their corresponding types.

| Service Operations | Parameters | Types |
|---|---|---|
| Visit Preparation Receive | PatientID | Primitive |
| Visit Preparation Reply | VisitNote | Complex |

### 4.1 Description mismatch analysis

Each BPEL process is a web service and needs a WSDL document. A WSDL document describes a web service using some elements including <types>, <message>, <portType>, and <binding>.

A client will usually invoke an operation on the BPEL process to start it. With the BPEL process WSDL, we specify the interface for this operation. We also specify all message types, operations, and port types a BPEL process offers to other partners. However, the source of description mismatches come from operation message number and their data types.

The <types> element encloses data type definitions that are relevant for the exchanged messages. For maximum interoperability and platform neutrality; WSDL uses XML schema syntax to define data types. The XML schema distinguishes between primitive, derived, and complex data types[39].

**Primitive data types:** Primitive data types are those that are not defined in terms of other data types. Currently, XML schema defines 19 primitive data types [40].

**Derived data types:** Derived data types are those that are defined in terms of other data types. There are three kinds of derivation: by restriction, by list, and by union. The XML Schema has 25 derived types build-in[40].

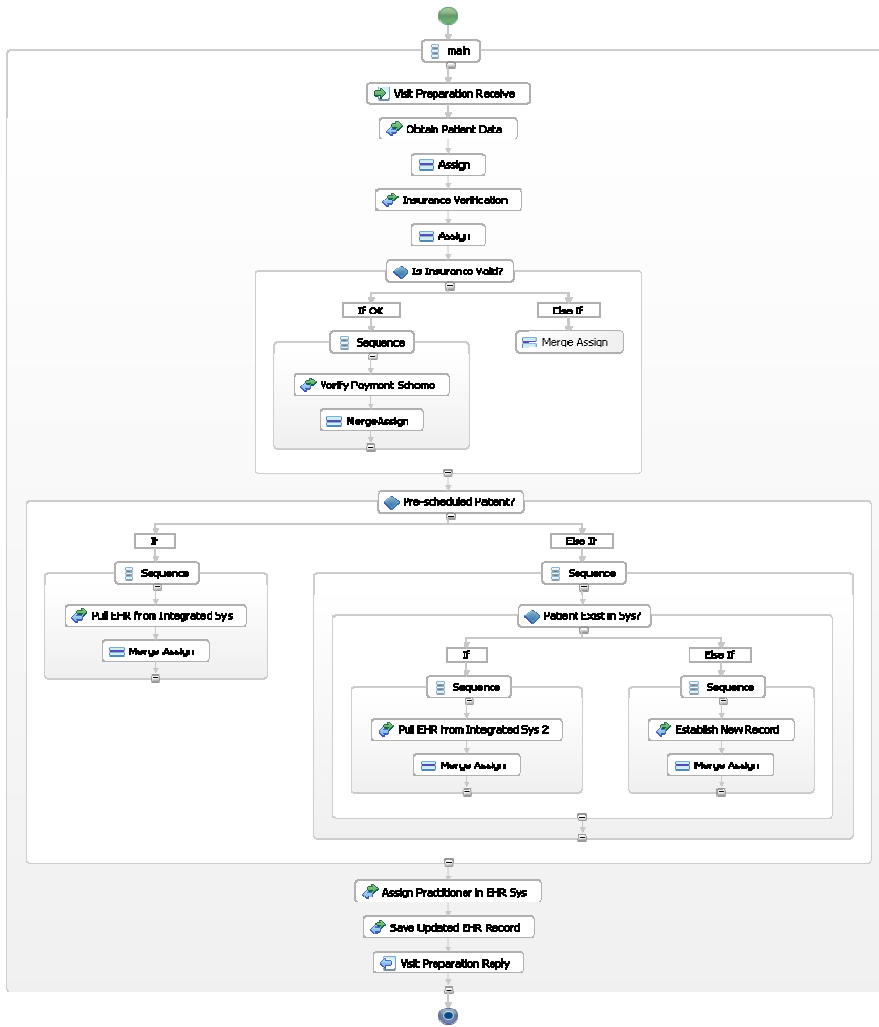

Fig. 2: *Visit Preparation* BPEL process.

**Complex data types:** XML Schema provides a flexible and powerful mechanism for building complex data structures from its simple data types. Data structures can be created via making a sequence of elements and attributes. Additionally, user defined types can be extended in order to create even more complex types. XML Schema has three compositor elements that allow constructing complex data types from simpler ones: sequence, choice, and all. The behavior of these compositor elements is defined as follows:

- **Sequence**: All the complex type fields must be present in the exact order they are specified in the type definition.

- **All**: All of the complex type's fields must be present but can be in any order.

- **Choice**: Only one of the elements in the structure can be placed in the message.

Data types are different from each other on the basis of their usage and mismatch-proneness. This is due to the fact that some types could be covered or cast by some other data types. In practice, for example, Strings could be used much frequently across an application than many other types. A BEPL workflow matching only against Strings may be much more reusable than one matching against an application-specific data type. More exactly, String covers all data types (that is 19 primitive data types plus 25 derived types) and its $w$ is the value of 44, hence its coverage weight become the value of 1 (i.e. 44/44). In a similar approach, the Boolean and Float data types cover 1 and 15 data types, so their coverage weight become the value of 0.02 (i.e. 1/44) and 0.34 (i.e. 15/44) respectively. Based on the mentioned hypothesis, description mismatch probability calculation is done as follows:

Let.

- *MP*: refers to match probability.

- *MMP*: refers to mismatch probability.

- $MP_P$: refers to match probability of input/output parameters data types.
- $MP_{P_k}$: refers to match probability of input/output parameters data types of operation $k$.
- $MP_{CD}$: refers to match probability of a complex data type.
- $MMP_{CD}$: refers to mismatch probability of a complex data type.
- $MP_{SD}$: refers to match probability of service description.
- $MMP_{SD}$: refers to mismatch probability of service description.
- $d$: refers to the total number of defined or built-in primitive and derived type in XML schema. Thus $d$ has the constant value of 44.
- $w_j$: refers to the number of data types that are covered by each data type.
- $l$: refers to the number of primitive, derived, or complex data types of an operation.
- $l_k$: refers to the number of primitive, derived, or complex data types of operation $k$ parameters.
- $n$: refers to the number of simple or even another complex type within Sequence, Choice, or All element.
- $O_k$: refers to service operation $k$.
- $m$: refers to the number of service operations.

The equation 1 calculates match probability of both input and output parameters data types.

$$MP_P = \begin{cases} \prod_{j=1}^{l}\left(\dfrac{w_j}{d}\right)_j & \textit{primitive type} \\[2ex] \prod_{j=1}^{l}\left(\dfrac{w_j}{d}\right)_j & \textit{Derived type} \\[2ex] \prod_{j=1}^{l}\left(MP_{CD_j}\right) & \textit{Compelex type} \end{cases} \qquad (1)$$

As specified earlier, twenty-five derived types are defined within the specification itself, and further derived types can be defined by users in their own schemas. We refer to those derived types out of the 25 built-in ones as complex types. As complex data types include some primitive data types with *Sequence* or *Choice* behavior, for example, hence their match/mismatch probability must be calculated in terms of these behaviors, equation 2.

$$MP_{CD_j} = \begin{cases} \left(\dfrac{1}{n!}\right)\times\left(\prod_{i=1}^{n}\left(\dfrac{w_j}{d}\right)_i\right) & \textit{Sequence} \\[2ex] \left(\dfrac{1}{2^n}\right)\times\left(\prod_{i=1}^{n}\left(\dfrac{w_j}{d}\right)_i\right) & \textit{Choice} \\[2ex] \left(\prod_{i=1}^{n}\left(\dfrac{w_j}{d}\right)_i\right) & \textit{All} \end{cases} \qquad (2)$$

The events of occurring one of the possible data types for each of the input/output parameters are *independent events*. Based on probability theory, assuming independence (that is the mismatch of either parameter data type is not influenced by the success or failure of the other parameter data type), the service description mismatch probability can be calculated through probability calculation of the description not matching, or the matching of the service description (equation 3) and then, the probability of service description mismatch is simply 1 (or 100%) minus the match probability, equation 4.

$$MP_{SD} = \left(\prod_{k=1}^{m}\left(\frac{1}{l_k}\times MP_{P_k}\right)_{O_k}\right) \qquad (3)$$

$$MMP_{SD} = \left(1 - \left(\prod_{k=1}^{m}\left(\frac{1}{l_k}\times MP_{P_k}\right)_{O_k}\right)\right) \qquad (4)$$

This question may arise that "*Did the metric take the number of parameters as an important source of mismatch into account?*" The answer is positive. We accept that the more parameters lead to the more description mismatch probability and the $1/l_k$ coefficient covers this issue.

Now, based on the provided information about the scenario, we are able to calculate its description match probability. According to the table 1, PatientID is String data Type, while VisitInfo is a complex type with sequence behavior in it. VisitInfo consists of three primitive types including ID: String, VisitDate: String, and Medications: String. Finally, $MP_{SD}$ is calculated as follows:

$$MP_{PatientID} = \frac{44}{44} = 1 \qquad MP_{VisitNote} = \frac{1}{3!} \times \left( \prod_1^3 \frac{44}{44} \right) = 0.16 \qquad MP_{SD} = \frac{1}{2}(1 \times 0.16) = 0.08$$

### 4.2 Logic mismatch analysis

In addition to description mismatch concept, a BPEL process cannot be reused in potential contexts for the sake of disparate business rule which lead to disparate structure and logic i.e. logic mismatch. For the purpose of logic mismatch probability calculation, firstly we have to calculate match probability of each structured activities within a composite service. Each BPEL process consists of steps. Each step is called an activity. BPEL activities divided into Basic and Structured categories[41]. Basic activities including `<invoke>`, `<receive>`, `<reply>`, `<assign>` are used for common tasks such as invoking a web service or manipulating data. Structured activities are used for arranging the structure of BPEL process. Structured activities can contain both basic and structured activities in order to implement complex business processes. The important and common use structured constructs are `<sequence>`, `<flow>`, `<switch>`, `<pick>`, and `<while>`.The following subsections explore how match probability of BPEL process constructs is calculated.

#### 4.2.1 *<Sequence> match probability*

A `<sequence>` activity is used to define activities that need to be performed in a sequential order. The match probability (MP) of `<sequence>` activity is calculated as follows:

$$\text{MP}_i = \left( \frac{2^n - 1}{n! + 2^n - 2} \right) \qquad (5)$$

- *n:* is the number of activities within a sequence.
- *i:* refers to `<sequence>` construct.

In sequence construct, firstly we have to define when a composite service can be matched with potential solution without tough efforts. In our perspective, this can be achieved when the arrangement of activities does not change. To illustrate the point consider the sample BPEL processes that have been depicted in figure 3(a) and 3(b). Suppose an order-to-cash business process that can occur according to two distinct flows including payment-after-shipment or shipment-after-payment based on the context business rule. Provided the former flow, the total number of match and mismatch flow are the value of 7 and 12, respectively. In fact, if the arrangement of activities does not change, we could reuse the provided composite web service with minimal change. Such a change can also be a kind of omission for some of activities. In fact, matchable control-flows are those that could be realized with minimal modifications, while the other permutations cannot be realized straightforwardly.

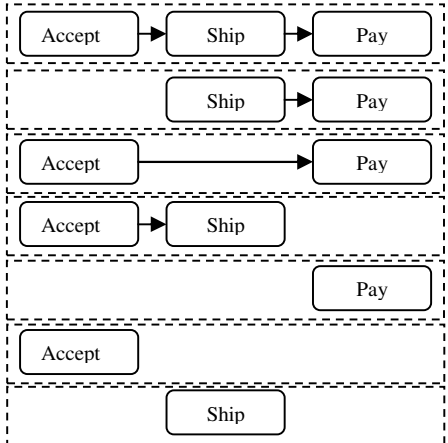

Fig. 3(a): Matchable control-flows of a given web process.

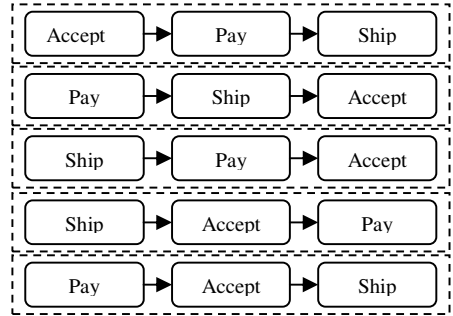

Fig. 3(b): Unmatchable control-flows of a given web process.

Therefore, the match probability is directly dependent on the different ways of arranging the activities in a sequence i.e. permutations. Thus, in case of 'n' activities, we had n! in addition to the number of omission cases (i.e. $2^n$-1) as a total number of sample space. Since there is one permutation in both n! and $2^n$-1, we have to subtracted them by the value of 1.

The numerator of equation 5 is the number of matches that are omission cases in addition to the unchanged case that is $2^n$-1. Based on the probability theory, the match probability of a service with one *sequence* activity computed via equation 5.

However, there is an exception when considering `<sequence>` activity in a BPEL code. It is expected that the `<sequence>` activity which is used upon for synchronous invocation of web service operations is not taken into account, since the permutation is not possible.

### 4.2.2 *<Switch> match probability*

The `<switch>` activity is used to express conditional behavior. It consists of one or more conditional branches defined by `<case>` elements, followed by an optional `<otherwise>` element. The case branches of the switch are considered in alphabetical order. The match probability of `<switch>` activity is calculated as follows:

$$MP_i = \left( \frac{1}{2^n} \right) \qquad (6)$$

- *n:* is the number of conditions.
- *i:* refers to `<switch>` construct.

The `<switch>` activity consists of an ordered list of conditions specified by a `<case>` element followed by one optional `<otherwise>` element. *Switch* activity can be changed based on its conditions. That is, each condition in a *switch* activity can or cannot be changed according to business rule. We compute the contingency of this behavior via the power set of the number of conditions, i.e. $2^n$. Therefore, the match probability involved in a *switch* construct is insignificant, since the match case is just the one in which none of the conditions have been changed.

### 4.2.3 *<Pick> match probability*

The `<pick>` activity is used to wait for the occurrence of an event and then perform an activity associated with the event. The match probability of `<pick>` activity is computed as follows:

$$MP_i = \left( \frac{2^n - 1}{2^{n+1} - 2} \right) = \frac{1}{2} \qquad (7)$$

- *n:* is the number of events which are cached through `<onAlarm>` and `<onMessage>` activities.
- *i:* refers to `<pick>` construct.

For the purpose of calculating match probability of `<pick>` activity we have to enumerate the number of match and mismatch cases. In *pick* construct the numbers of match cases are those in which the events conditions such as messages have not been changed. Moreover, deleted cases in which some of events removed regarding the business rules can also be contemplated as match cases. This is due to the fact that, composite service with such modifications can be reused without though efforts. In this matter, the total number of both match and mismatch cases are twice the match cases. Thus the result gets the value of ½.

### 4.2.4 *<Flow> match probability*

The `<flow>` activity provides concurrent execution of enclosed activities. The *flow* activity also allows the synchronization of activities within the *flow*. To do so, a *link* construct (`<Link>`) specifies a dependency between a source and target activity.

The match probability of *flow* activity is one, when the activities within a *flow* are processed on the condition that the order of execution is not defined. In this matter, the match cases are just the omission cases that are $2^n$-1. Since the numbers of unmatched cases are zero, so the match probability takes the value of one. This means, different versions of a same business rule does not affect the flow control of `<flow>` construct.

However, if we define any dependency between activities using `<Link>` construct, the match probability is dependent upon the different ways of arranging the activities in a *flow* construct. Since in dependency behavior activities rely on each other, hence unlike *sequence* construct omission cases are not acceptable. Therefore, the match probability of *flow* is calculated as follows:

$$\mathrm{MP}_i = \begin{cases} \dfrac{2^n - 1}{2^n - 1} = 1 & \text{The flow construct with concurrency behavior.} \\[2mm] \dfrac{1}{n!} & \text{The flow construct with dependency behavior.} \end{cases} \tag{8}$$

- *i:* refers to `<flow>` construct.
- *n:* is the number of activities within a flow.

### 4.2.5  *<While> match probability*

A `<while>` construct is used to define an iterative activity. The iterative activity is performed until the specified Boolean condition no longer holds true. The match probability of `<while>` construct has the value of 0.5, as the match can be occurred at Boolean condition of the activity:

$$\mathrm{MP}_i = \frac{1}{2} \tag{9}$$

- *i:* refers to `<while>` construct.

### 4.2.6  *Logic mismatch analysis*

In previous subsections, the match probability of all kinds of structured constructs was calculated. Since a BPEL process may have more than one of them at the same time, it is expected to compute the match probability in case of more than one structured activities in a BPEL process. Thus, logic match probability of a BPEL process with any number of structured constructs is calculated through equation 10.

$$\mathrm{MP}_{SL} = \left( \prod_i^n MP_i \right) \tag{10}$$

Where

- *i:* refers to specific structured construct.
- *n:* refers to the total number of structured constructs within a certain BPEL process.
- $MP_{SL}$: refers to the Service Logic (SL) match probability.
  Now, we are going to apply logic match probability equation to the under-discussion scenario (Fig. 2) as follows:

$$MP_{SL} = \left( \frac{2^6 - 1}{6! + 2^6 - 2} \right) \times \left( \frac{1}{2^2} \times \frac{1}{2^2} \times \frac{1}{2^2} \times \frac{1}{1!} \times \frac{1}{1!} \times \frac{1}{1!} \times \frac{1}{1!} \times \frac{1}{1!} \right) = 0.08$$

Logic mismatch probability of a composite service with any number of structured constructs is calculated via equation 11.

$$\mathrm{MMP}_{SL} = \left( 1 - \left( \prod_i^n MP_i \right) \right) \tag{11}$$

Where

- $MMP_{SL}$: refers to Service Logic (SL) mismatch probability.

### 4.3  *BPEL process reusability analysis*

Now, we are able to analyze the BPEL process reusability. In this regard, firstly we have to calculate a BPEL process total mismatch probability that is both description and logic mismatch probability. A BPEL process total mismatch probability is as follows:

$$\text{MMP}_\text{S} = \left(1 - \left[\left(\text{MP}_\text{SD}\right) \times \left(\text{MP}_\text{SL}\right)\right]\right) \quad (12)$$

Where

- $MMP_S$: refers to mismatch probability of a BPEL process.
- $MP_{SL}$: refers to the Service Logic (SL) match probability.
- $MP_{SD}$: refers to match probability of service description.
  Regarding to equation 12, the total mismatch probability for the scenario is calculated as follows:

$$\text{MMP}_\text{S} = (1 - (\text{MP}_\text{SD} \times \text{MP}_\text{SL})) = 1 - 0.006 = 0.994$$

Based on the calculation of BPEL process total mismatch probability, BPEL process reusability can be computed as follows:

$$R_p = R_c \times \left(1 - MMP_S\right) \quad (13)$$

Where

- $R_c$: refers to the numbers a BPEL process reused in current context.
- $R_P$: refers to potential reusability of a BPEL process in potential contexts.
  Equation 13 computes the potential reusability of a given BPEL process with respect to the mismatch probability of a BPEL process in the prospect contexts.
  Now, we have to apply the equation 13 to the scenario. Potential reusability of Visit Preparation BPEL process (Fig. 2) calculated with the assumption that the current reusability of the service is the value of 1 (i.e. $R_{C=}$1).

$$R_p = 1 \times \left[1 - MMP_S\right] = 0.006$$

## 5. Validation of the Reusability Metric

### 5.1. Theoretical Validation

In this section, we aim to theoretically validate the proposed metrics using the reusability properties that are proposed in [26] which themselves have been inspired from the property-based framework of Briand et al.[42]. This framework is a mathematical generic framework that introduces some intuitive properties for salient concepts of software engineering such as complexity, coupling, cohesion, and size, through which researchers and practitioners could analyze and validate the theoretical grounds of their measures irrespective of a specific development paradigm. Therefore, we examine the proposed reusability metrics against four reusability properties [26] including Non-negativity, Null value, Monotonicity, and Merging of modules.

- **Property 1**: *Non-negativity*: This property is satisfied since for a given BPEL process the value of reusability metric is equal to zero when the service does not have any consumer or some positive value representing the potential reusability value of a BPEL process. As a result, it will never be negative under any circumstances.
- **Property 2**: *Null value*: This property is also satisfied since in case of no consumers the $R_c$ gets the value of zero, hence the reusability of a BPEL process (i.e. $R_p$) gets zero too. So, the reusability will be null in case there are no consumers.
- **Property 3**: *Monotonicity*: This property is satisfied because the reusability value of a BPEL process cannot be decreased when the number of consumers increases. To be more specific, by increasing the number of consumers the $R_c$ becomes greater and subsequently the $R_p$ gets the higher value.
- **Property 4**: *Merging of modules*: This property is also satisfied since the reusability of a BPEL process obtained by merging two processes is not greater than the sum of reusability of the two original ones. This is due to the fact that the obtained process is more granular so that has less potential reusability.

As proposed potential reusability metrics adhere to all the prescribed properties for the corresponding attribute, hence it can be considered as a valid characterization of potential reusability of a BPEL process.

### 5.2. Empirical Validation

Although the proposed metrics are theoretically valid, this does not verify the predictive power of the metric in terms of an empirical relationship between the metric and the potential reusability they meant to predict. This section presents an experiment that authors have conducted to empirically validate the proposed metrics. Additionally, inferential and interpretation analysis of the metrics and their relation to the structural elements of the experimental material through statistical diagrams are explained. Meanwhile, power analysis of the hypothesis test and the estimation of the associated confidence interval are computed accordingly.

For the experiment to be successful it needs to be wisely constructed and executed. Therefore, we have followed some suggestions, provided by Perry et al. [43] and Mendonca et al. [44], about the structure and the components of a suitable empirical study. To perform an experiment, several steps have to be taken in a certain order. An experiment can be divided into the following main activities [44]: goals of the study, hypotheses, experimental protocol, threats to validity, data analysis and presentation, results and conclusions. In the remainder of this section we will explain how these activities have been performed.

### 5.2.1. Goal of the study

The main goal of this study is "Analyzing the proposed reusability metrics to evaluate their predictive capability with respect to the design-level estimation of the potential reusability of the BPEL processes."

### 5.2.2. Hypothesis Formulation

Hypotheses are essential as they state the research questions in a semi-formal way. We present our hypothesis in two levels of abstraction as follows:

**Abstract Hypothesis:** "The potential reusability ($R_P$) metric is a good predictor and accurate metric to evaluate the potential reusability of composite services."

**Concrete Hypothesis:** "There is a significant correlation between the potential reusability ($R_P$) metric values and the expert's rating of the potential reusability of a set of processes as our experimental material."

### 5.2.3. Experimental Protocol

Having formulated the hypotheses, the design of the experiment occurred precisely according to the guidelines. An experimental design is a detailed plan for data collection and the other experimental tasks that will be used to test the hypotheses. This phase also explains how the experiment was conducted and has several components that are fully described in order to provide useful information for future replications.

#### 5.2.3.1. Variable selection

Typically, in empirical studies there are two kinds of variables including dependent and independent that their cause and effect should be evaluated by testing the hypothesis with appropriate techniques. In this research they are as follows:

- The independent variable is the structure of BPEL processes.
- The dependent variable is the potential reusability of processes which varies when the structure of BPEL processes changes.

#### 5.2.3.2. Expert selection

The experts we selected were students of the Electrical and Computer Engineering Faculty enrolled in the Master and PhD program in Computer Engineering at Shahid Beheshti University, Tehran, Iran. Since the desired population for the study was rare and very difficult to locate, they were selected based on purposive sampling [51]. Twenty experts were selected according to the evaluations of their lecturers. Half of the participants were male in the 24–50 age-groups and the rest were female in the 23–29 age-groups. Moreover, all participants were volunteers who had interest in software engineering research. Some of them had extensive industrial experience in several areas, but none had experience with business process management systems. By the time the experiment was done, all the students had taken a 50 hour course on ULS (Ultra Large-Scale Systems) with emphasis on service-oriented systems, business processes modelling, and service composition, therefore, gained experience in the design and development of services specially business services. To enhance their knowledge about service modelling, a group-based training session was carried out before doing experiment. This session consisted of an introduction to BPEL language, its constructs, and the quality attributes of a BPEL process.

#### 5.2.3.3. Experiment design

The objects supposed to be rated by experts were process-based services graphically designed with the Eclipse BPEL designer. These objects were collected by the members of ASER group. The material consisted of 70 professionally designed composite services of the different universe of discourses such as university, core-banking, travel agency, etc. with different structural characteristics and degrees of complexity. The participants were told how to carry out the experiment with exactly the same set of 70 pre-designed composite services. In order to make the experience and knowledge of the participants more comparable, we made 10 groups out of the 20 participants. Table 2, denotes participant groups and associated profiles.

The participants could use unlimited time to make a consensus based on their judgments. Our key aim to give them unlimited time was based on the fact that we did not intend to rush their consensus. During the workshop they felt free to discuss enough to make a consensus. However, since the group's experiences were comparable they finished rating after 3 hours. In order to gather more precise rating, authors decided to design a professional questionnaire consisting the following questions that each of which investigates one aspect of service reusability. The first two questions are related to description, whereas the others are concerned with service logic.

- To what extent is the composite service interface granular?
- To what extent is the composite service interface complex?
- To what extent is the composite service logic granular?
- To what extent is the composite service logic complex?
- To what extent is the composite service abstract?
- To what extent is the composite service context-independent (decouple from context)?

These aspects including complexity, coupling (context-dependency), granularity, abstraction, etc. are, in fact, the inherent indicators of service reusability [16][45]. Although the rating score does not directly indicate the reusability of a service, the above-mentioned characteristics that were considered in the review tend to affect the service reusability [15]. Therefore, the rating score reflects the reusability of a service, and also the quality factors that have impact on the reusability. The reason of gathering information in an indirect manner is according to statistics since direct questionnaire may lead to bias results. Moreover, we believe that the respondents could not rate reusability attribute intuitively.

The independent variable was measured using the $R_P$ metric formulated in section 4. The dependent variable was measured according to average of the expert's ratings for each question that has been rated by the participants from 1 to 10. These values are contemplated as to be on an interval scale for the analysis, since the difference between two values is meaningful. Additionally, the questions are related to the reusability in the opposite manner except the last one that has positive effect on it. It means when the participants rate this question, the small numbers indicate worse situation in the viewpoint of them.

Table 2. Participant groups and associated profiles.

| Group Index | No of Experts | Profiles |
|---|---|---|
| 1 | 2 | Master students in software engineering |
| 2 | 2 | Master students in software engineering |
| 3 | 3 | Master students in IT |
| 4 | 2 | Master students in IT |
| 5 | 2 | Master and PhD Students in software engineering |
| 6 | 2 | Software engineering practitioners |
| 7 | 2 | Master Students in software engineering |
| 8 | 3 | Master Students in software engineering |
| 9 | 1 | PhD student in software engineering |
| 10 | 1 | Professor in software engineering |

### 5.2.4. Analysis of the Results

Since the experts rated services using a numerical scale, we have selected quantitative analysis to draw conclusions from the data. The qualitative analysis was done in conjunction with a number of statistical analyses.

### 5.2.4.1. Descriptive Statistics

The descriptive statistics are presented in Tables 3 for each independent and dependent variable.

Table 3: Descriptive Statistics.

| | | Mean | Std. Deviation | Variance | N |
|---|---|---|---|---|---|
| Group | 1 | .71 | .15 | .02 | 70 |
| | 2 | .69 | .15 | .02 | 70 |
| | 3 | .72 | .14 | .02 | 70 |

| | | | | |
|---|---|---|---|---|
| 4 | .78 | .13 | .02 | 70 |
| 5 | .60 | .16 | .03 | 70 |
| 6 | .73 | .13 | .02 | 70 |
| 7 | .71 | .14 | .02 | 70 |
| 8 | .65 | .10 | .10 | 70 |
| 9 | .73 | .13 | .02 | 70 |
| 10 | .70 | .13 | .02 | 70 |
| $R_P$ | | .30 | .34 | .12 | 70 |

### 5.2.4.2. Hypothesis Testing

In this section, it is investigated if any correlation exists between experts' ratings and the proposed $R_P$ metric value. The first step in correlation analysis is to ascertain whether the distribution of the data is normal, so the Kolmogorov-Smirnov test was applied. As it is obtained that the distribution was not normal, authors decided to use a non-parametrical statistical test, namely the Spearman Correlation [46] with the level of significance of $\alpha 1=0.01$ which indicates the probability of rejecting the null hypothesis when it is certain (type I error). The Spearman $r_S$ is a non-parametric statistic used to show the predictive relationship between the two variables which are expressed as ranks (the ordinal level of measurement). The correlation coefficient is a measure of the ability of one variable to predict the value of another one. This does not imply the existence of any causal relationship between the two variables. Indeed, it just indicates a predictive relationship. Using Spearman's correlation coefficient, the $R_P$ metric was correlated separately to the different experts' rates of reusability. In the experiment, the null hypothesis was:

**$H_0$:** "there is no significant correlation between the $R_P$ metric and the experts' rating of potential reusability".

**$H_1$:** "there is a significant correlation between the $R_P$ metric and the experts' rating of potential reusability".

The probability that the null hypothesis would be erroneously rejected was controlled with confidence level $\alpha 1=0.01$. An analysis performed on the collected data and Table 4 denotes the summary of results describing the Spearman correlation coefficient between experts' ratings and the rates given by the $R_P$ metric in the experiment. The absolute value of the correlation coefficient indicates the strength, with larger absolute values indicating stronger relationships. The significance level (or p-value) is the probability of obtaining results as extreme as the one observed. Based on the results which are denoted in Table 4 and taking $\alpha 1$ into consideration, correlation is significant at the 0.01 level for 100% of the experiment; therefore, the null hypothesis was rejected with 99% confidence.

Table 4. The Spearman correlation coefficient between experts' ratings (the academics) and the values given by the $R_P$ metric.

| | | $r_s$ | $\alpha 1$ | N |
|---|---|---|---|---|
| | 1 | .530 | Reject H0 | 70 |
| | 2 | .585 | Reject H0 | 70 |
| | 3 | .580 | Reject H0 | 70 |
| | 4 | .610 | Reject H0 | 70 |
| Group Index | 5 | .538 | Reject H0 | 70 |
| | 6 | .561 | Reject H0 | 70 |
| | 7 | .521 | Reject H0 | 70 |
| | 8 | .479 | Reject H0 | 70 |
| | 9 | .589 | Reject H0 | 70 |
| | 10 | .600 | Reject H0 | 70 |
| Average | | .635 | Reject H0 | 70 |

### 5.2.4.3. Power Analysis of the Hypothesis Test

A test without sufficient statistical power will not produce enough information to convince the researcher regarding the acceptance or rejection of the null hypothesis [47]. Since we are using relatively medium samples, authors provide details of power analysis in this section to interpret the results. The power of the nonparametric tests is determined by appropriate parametric tests [47]. Therefore, we have used Pearson's $r$ [48] for the Spearman Rank Correlation.

Given: i) observation size of 70 from 10 group of participants that produce 700 data points; ii) level of $\alpha= 0.01$; and iii) the effect size 0.6; the achieved power is 0.81 that is considerably above the accepted norms [47] in software engineering studies. However, by the effect sizes lower than 0.6 the power would dramatically shrinks. Thereby suggesting that the statistical tests could produce a Type II error [47] (i.e. fail to reject a null hypothesis when it is in fact false) when the test do not exhibit large effect sizes.

### 5.2.4.4. Estimation of the $R_P$ confidence interval

Metrics without any thresholds or intervals for measurement values may not be effectively used [22]. Therefore, we decided to calculate the confidence intervals of the metrics. We calculated the confidence intervals of $R_P$ in terms of the reusable services using the rating scores. First, we assumed that the reusability of services that satisfy average rating score higher than 0.75 is high. Second, by an interval estimation based on the Wilcoxon signed-rank (non-parametric) test statistic [49], we calculated confidence intervals with confidence coefficient 95% for $R_P$ using all services that satisfy the condition among all samples. The number of such high rated services was 29. The average of measurement values of all samples is 0.29, and the number of services whose measurement values are in the confidence interval is 8. The lower confidence limit is 0.375 and upper confidence limit is 0.625.

Among all samples, the percentage of services whose values of $R_P$ are in the confidence interval is only 11%. This means that the confidence interval of $R_P$ is not a general range where every service corresponds. Therefore, it is evident that $R_P$ with its confidence interval can be effectively used for measuring the potential reusability of services.

From the obtained confidence interval, users of the proposed metric can judge the potential reusability of a service at hand. From statistical point of view, if the potential reusability of a service were in the confidence interval, then by 95% it is possible that the experts will give it a rate higher than 0.75. Moreover, it might be possible that the potential reusability was not in the confidence interval or even higher than the upper limit, then it should be evaluated what causes undermine the other quality factors for the sake of reusability.

### 5.2.4.5. Descriptive statistics and interpretive analysis of $R_P$

In this subsection, the $R_P$ metric will be analysed from different viewpoints, thereby supporting the interpretation of the experimental results. In this regard, authors investigate the relationship between $R_P$ and granularity, logical constructs, and data structures. Such critical analyses based on the gathered data let us to interpret the behaviour of the introduced metrics.

Figure 4 shows the number of BPEL processes in each section among ten split sections with respect to measurement values of $R_P$. The frequency of services whose values of $R_P$ are lower than .2 and higher than .9 tends to be high.

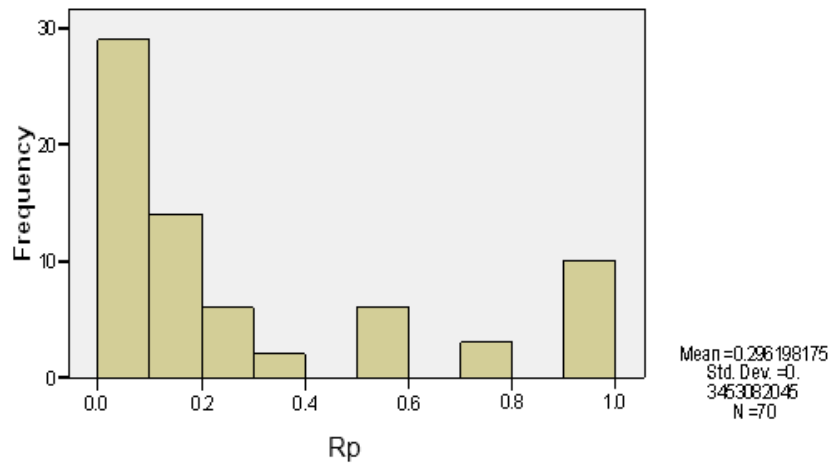

Fig. 4: The frequency of composite services with respect to the $R_P$ values.

### $R_P$ and Granularity:

Granularity is defined as how much functionality is exposed by a service. Service granularity and reusability has Inverse relationship [17] in which when the number of functionality (i.e. generality) decreases, the module cohesiveness becomes stronger and more cohesion leads to better reusability[19]. Figure 5 shows the average of potential reusability in each section among nine split sections with respect to granularity values of services. The average of potential reusability whose values of granularity are lower than 2 tends to be high. This means that services with too high granularity level are not reusable enough and this consistent with the average of expert's rating in the Fig 6.

***R<sub>P</sub> and Logical Constructs:***

To investigate the behaviour of $R_P$ in terms of logical constructs consider Fig. 7 that shows the average of potential reusability in each section among split sections with respect to number of logical constructs in the composite services. The average of potential reusability whose number of logical construct are lower than 8 tends to be high. This means that services with too high number of constructs are not reusable enough and this consistent with the average of expert's rating in Fig. 8. The number of logical constructs such as "*sequence*" as depicted in Fig. 9 should be lower to some extent as the high correlation between $R_P$ and number of construct in Table 5 testifies. In contrast the construct such as "*pick*", "*flow*", and "*while*" have no significant effect on the potential reusability as the low correlation between them confirms.

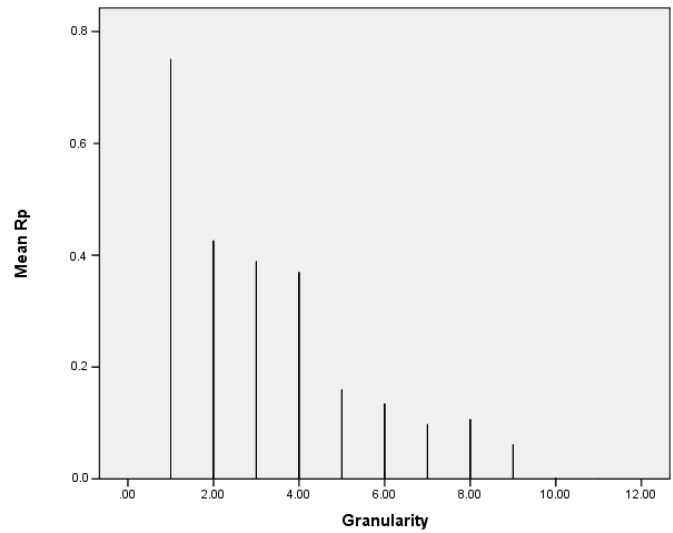

Fig. 5: The relationship between the mean reusability and granularity of composite services.

Table 5. The Spearman correlation coefficient between behavioral and structural construct of services and the values given by the $R_P$ metric.

| Flow | Pick | Switch | Sequence | Granularity | |
|---|---|---|---|---|---|
| .257(*) | .231 | .326(**) | .510(**) | .428(**) | Spearman's $R_P$ |
| Construct | Data Structure | String | While | | |
| .576(**) | .392(**) | .324(**) | .219 | | Spearman's $R_P$ |

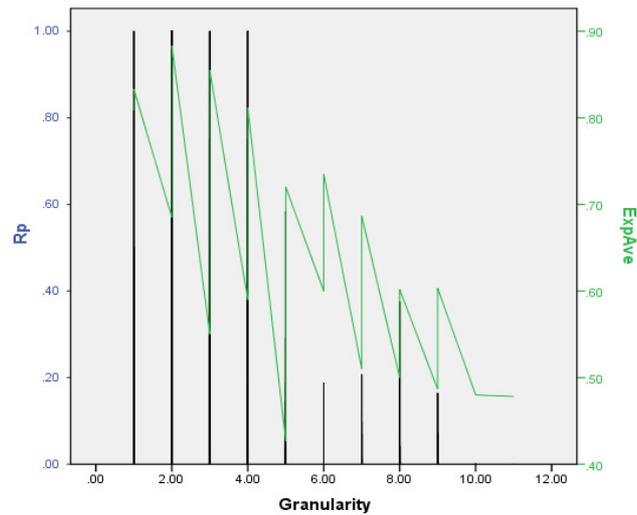

Fig. 6: The relationship between experts' rating on the reusability and granularity of composite services

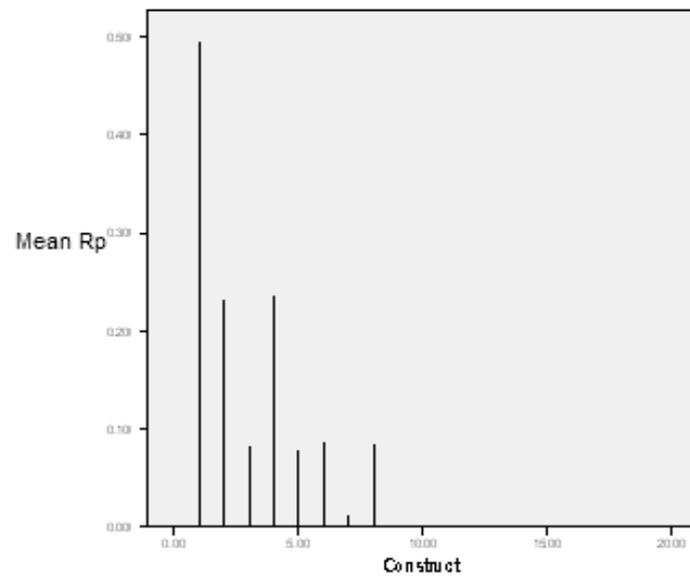

Fig. 7: The relationship between the mean reusability and the number of logical constructs.

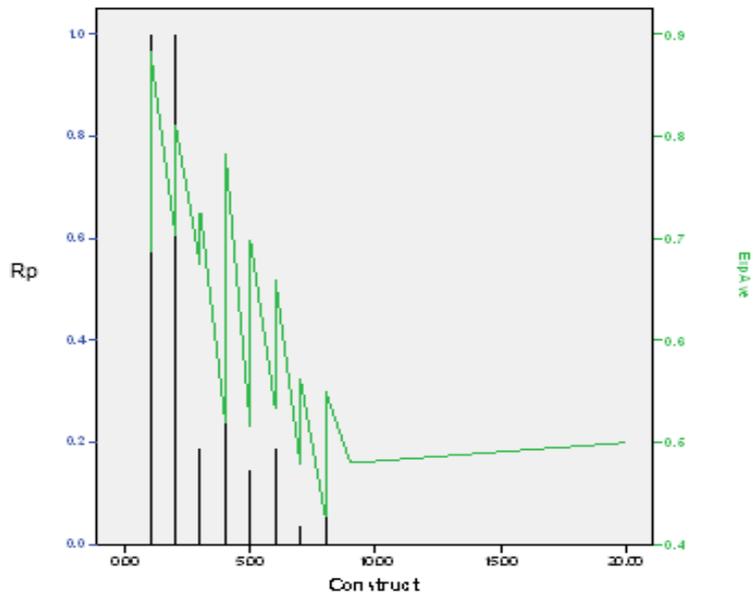

Fig. 8: The relationship between experts' rating on the reusability and the number of logical constructs.

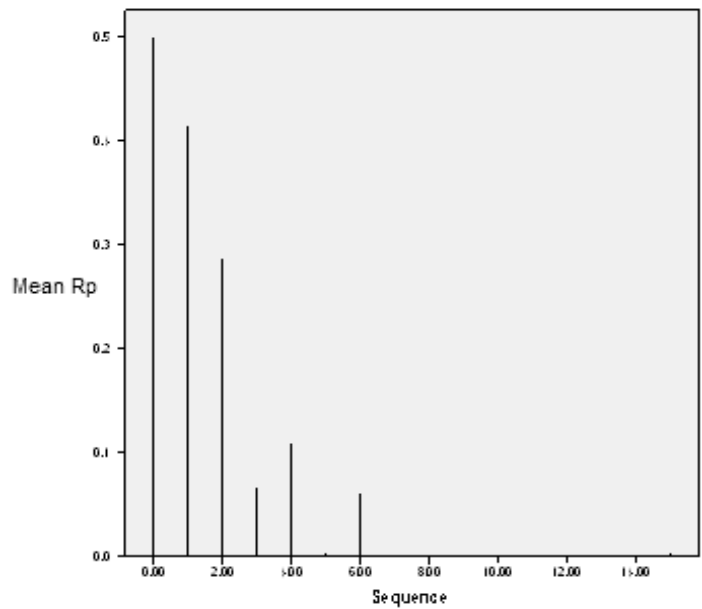

Fig. 9: The relationship between the mean reusability and the number of sequence constructs.

***RP and Data Structures:***

To study the behaviour of $R_P$ in terms of data structures consider Fig. 10 that presents the average of potential reusability in each section among split sections with respect to number of data structures in the services' signature. The average of potential reusability whose number of data structures are lower than 5 tends to be high. This means that services with too high number of data structures are not reusable enough and this consistent with the average of expert's rating in Fig. 11. The number of data structures should be lower to some extent as the high correlation between $R_P$ and number of data structures in Table 5 testifies.

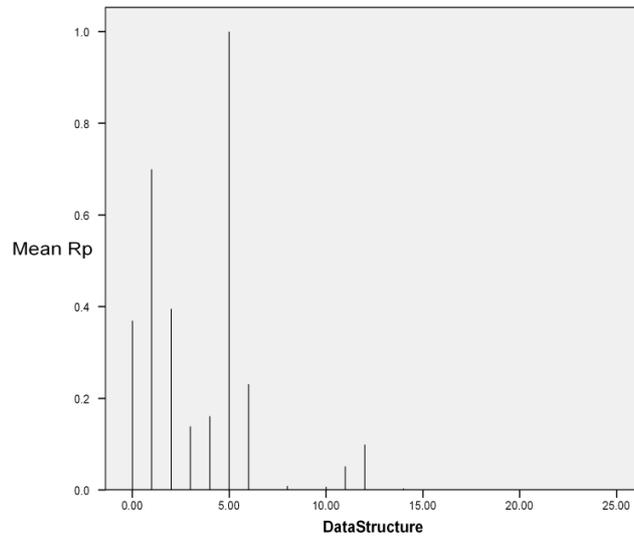

Fig. 10: The relationship between the mean reusability and the number of data structures.

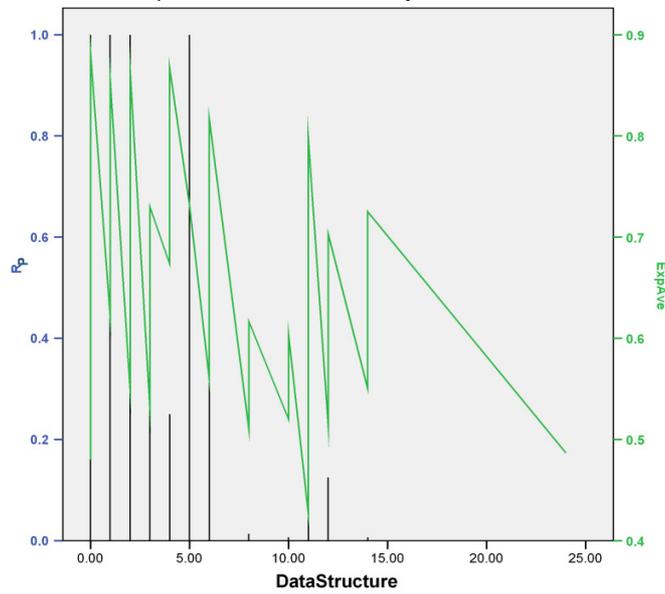

Fig. 11: The relationship between experts' rating on the composite service mean reusability and the No. of data structures.

### 5.2.5. Threats to Validity

Threats to validity may cast several effects on the interpretation of the experimental data. Thus, in this section we discussed various threats to validity in which they were categorized according to[43] and the way we attempted to alleviate them is as the way that Cardoso has conducted it in[50].

### 5.2.5.1. Construct validity

The independent variable that estimates the potential reusability of BPEL processes can be considered constructively valid because they are defined in a formal manner and also theoretically validated in Section 5.1.

*5.2.5.2. Internal validity*

We have considered different aspects that could threaten the internal validity of the study such as differences among experts, precision of experts' ratings, learning effects, fatigue effects, and experts' incentive.

- Differences among experts: We grouped the participant as described before; therefore error variance due to differences among participants was reduced. Additionally, purposive sampling [51] was employed to select participants with comparable knowledge and experience. Experts involved in this experiment had medium experience in service design and they attain minimum knowledge of BPEL process design, their ratings can be considered significant.
- Learning and fatigue effects: Since the experiment materials were in the different universe of discourses with different structural characteristics and degrees of complexity and the participants were in the small groups discussing on the issues to reach an appropriate consensus, it is very unlikely that any potential learning and fatigue affects the data.
- Anticipation effects: The participants were not told about the hypothesis that we wanted to test in order to ensure that expectations about specific levels of treatment did not influence their rates.

*5.2.5.3. External validity*

One threat to external validity is the number of participants that is limited to 20. This threat can limit the ability to generalize the results to settings outside the study.

- Experimental materials and Environment: The materials of our study are represented with BPEL and WSDL which both of them were standard and are utilized in industrial environment. Additionally, they were representative of real world BPEL processes in terms of size and complexity.

## 6. Discussion and future direction

In the experiment explained in section 5, authors came across with a number of cases in which the $R_p$ values were far from the experts' rating. More investigation shows by determining the weights for $MP_{sd}$ and $MP_{sl}$ the results will be more accurate and reliable. In fact, $MP_{sd}$ and $MP_{sl}$ seem to have different weights in constituting the $R_P$. In this regard, the correlation between $R_P$, $MP_{sd}$, $MP_{SL}$, Experiments Average (ExpAve), Experiments Average value for service logic (ExpLogic), and Experiments Average value for service description (ExpDes) were calculated. According to the obtained results, the correlation between ExpAve and ExpLogic (i.e. 0.968) is considerably more than the correlation between ExpAve and ExpDes (i.e. 0.508). In this matter, the behaviour of our metrics is a little bit different in which the correlation between $R_P$ and $MP_{sl}$ (i.e. 0.500) is close and a little less than the correlation between $R_P$ and $MP_{sd}$ (i.e. 0.628). Therefore, we have to define some weights for $MP_{sl}$ and $MP_{sd}$ by means of some methods particularly regression, which is part of our future work.

Additionally, some interesting results were deduced. For instance, the correlation between ExpDes and ExpLogic is insignificant (i.e. 0.323) which is somewhat in accompany with the correlation between $MP_{sd}$ and $MP_{sl}$. This means, the description and logic mismatch are the independent variables, thereby it is possible to simply multiply one by the other as it is applied in equation 12. Moreover, the correlation between $MP_{sl}$ and ExpLogic is noticeably significant (i.e. 0.800) and also the correlation value between $MP_{sd}$ and ExpDes is high (i.e. 0.574). These two correlations further confirm that the $MP_{sl}$ and $MP_{sd}$ respectively are the correct indicators of composite service logic and description match probability.

## 7. Conclusion

In this article, we proposed a metric for a BPEL process potential reusability analysis. We also reported results from an experiment designed to investigate its validity. The obtained results reveal that there exists a reliable statistical correlation between the proposed metric and the experts' judgments.

## Acknowledgements

Part of this study was carried out when the authors were in ASER group of Shahid Beheshti University and we would like to thank the involved computer science graduate students of Shahid Beheshti University and also Dr. Fereidoon Shams for his valuable supports. This work was also supported, in part, by Science Foundation Ireland grant 10/CE/I1855 to Lero - the Irish Software Engineering Research Centre.